# Lattice QCD and String Theory

**Julius Kuti**[*]

*University of California, San Diego*
E-mail: `jkuti@ucsd.edu`

Bosonic string formation in gauge theories is reviewed with particular attention to the confining flux in lattice QCD and its string theory description. Recent results on the Casimir energy of the ground state and the string excitation spectrum are analyzed in the Dirichlet string limit of large separation between static sources. The closed string-soliton (torelon) with electric flux winding around a compact dimension and the three-string with a Y-junction created by three static sources are also reviewed. It is shown that string spectra from lattice simulations are consistent with universal predictions of the leading operators from the derivative expansion of a Poincare invariant effective string Lagrangian with reparameterization symmetry. Important characterisitics of the confining flux, like stiffness and the related massive breather modes, are coded in operators with higher derivatives and their determination remains a difficult challenge for lattice gauge theory.

*XXIIIrd International Symposium on Lattice Field Theory*
*25-30 July 2005*
*Trinity College, Dublin, Ireland*

[*]This work was supported by the DOE, Grant No. DE-FG03-97ER40546.

# 1. Introduction

*Preamble*

In this review, I will identify some of the most important developments and challenges in the intriguing connection between lattice QCD and string theory. I kept the written version for the conference proceedings very close to the talk and its posted version at the conference web site. I hope that the expanded narrative will help the clarification of some important aspects of string formation including several questions and comments I received after the talk. I am thankful to my collaborators K.J. Juge, F. Maresca, C. Morningstar, and M. Peardon for some of the figures representing results before journal publications and left exactly in the form as they were shown in the talk. Keeping the size of the document within reasonable limits, I made every effort to preserve the resolution of the figures readable within the pdf format when magnification is used. The animation of string formation, which is embedded in Fig. 2, can be played within the document using Adobe Acrobat 6.0.

*Scope of the review*

When we probe the nature of quark confinement in simulations, our first goal is the high precision determination of the ground state energy and the excitation spectrum of a static quark–antiquark pair at large separation $R_{q\bar{q}}$. The microscopic understanding of the confining fuzz in the QCD vacuum of Fig. 1 remains a mystery but a legitimate question can be asked: Can the ground state energy and the excitation spectrum be described at large $R_{q\bar{q}}$ by the dynamics of collective string variables connecting the quark and the antiquark? The limit of large $R_{q\bar{q}}$ without string breaking only makes sense in the large *N* limit which can be studied in the quenched gauge sector without quark-antiquark vacuum loops. In this limit the opportunity for contact with string theory is rather obvious which explains the growing interest of string theorists in what lattice QCD can contribute to this exciting field. Perhaps the most basic question to ask is how the lattice excitation spectrum of Goldstone modes, associated with the confining flux of gauge theories, leads to geometric string variables on a two-dimensional world sheet with an effective Lagrangian which is Poincare invariant with reparameterization symmetry in non-critical physical dimensions. This will be the starting point of my report which will include the review of recent work on the Casimir energy of bosonic strings and the Y-shaped baryon string, the excitation spectrum of the bosonic Dirichlet string, and the closed string with unit winding number (string-soliton). I will develop an effective string theory description for the interpretation of the simulations which should provide the appropriate framework for future investigations in the field.

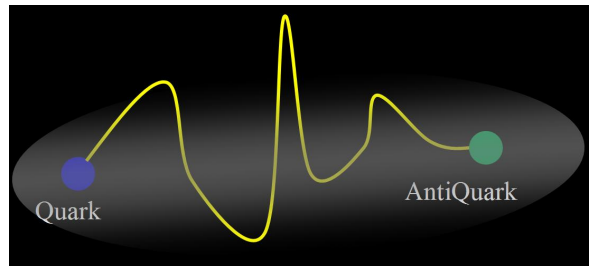

**Figure 1:** Strings and quark confinement

*Early history*

This review would not be complete without giving credit to early pioneering work from the long and curious history of the QCD string. My selection is bound to remain incomplete and only a more comprehensive survey of the field could do full justice to all the valuable work. After the first classic study of Nielsen and Olesen [1] on string formation in the Abelian Higgs model and its connection with the Nambu-Goto (NG) string, Polyakov reported a field theory model with microscopic confinement mechanism and string formation in the monopole dominated euclidean vacuum [2]. The first comprehensive study of large Wilson surfaces in QCD was published by Lüscher, Symanzik, and Weisz [3]. Shortly afterwards, Lüscher provided the first quantitative argument for the universal nature of the string Casimir energy [4]. The phenomenology of gluon excitations between static color sources was worked out by Hasenfratz et al. [5]. Particularly important was the first pioneering lattice study of gluon excitations between static color sources by Michael and Perantonis [6]. The effective Lagrangian of non-critical strings was developed by Polchinski and Strominger [7] providing a new framework for the discussion of string spectra in lattice QCD and circumventing the quantization problems of the Nambu-Goto string.

*Recent developments*

With new lattice technologies, comprehensive simulations of string spectra was brought to a new level of reach and accuracy by Juge et al. [8]. The lattice exploration of string behavior in SU(N) Yang-Mills gauge theories in the the large N limit was initiated by Teper [9]. Gliozzi and his collaborator presented a very useful series of studies in three-dimensional Z(2) lattice gauge theory [10]. They studied the effects of string imprints on the free energy of large Wilson loops while Hoppe and Münster calculated the string tension in the loop expansion starting from the classical string solution [11]. Some important recent results on the excitation spectrum of the Z(2) gauge model will be discussed in this review [12, 13, 14].

*Related topics*

I had no time in the talk to discuss recent results on the large N study of string formation which is the topic of Teper's contribution [15]. An important new approach to large N lattice QCD is addressed in the plenary talk of Neuberger [16] while for the exploration of the ADS/CFT connection I refer you to Brower's recent work [17]. String breaking in unquenched QCD is discussed in Bali's talk [18] and on-lattice k-strings are discussed in Gliozzi's talk [19]. Unfortunately, other interesting topics, which include 't Hooft loops, deconfinement and the string Hagedorn temperature, had to be omitted as well.

## 2. String formation in field theory

### 2.1 Animation in real time

Before I review the origin of string formation in field theory, it is helpful to illustrate visually what happens when static color sources are pulled apart to large separation in a confining theory. In Fig. 2 some excitations of the gauge field are shown in three-dimensional Z(2) lattice gauge theory.

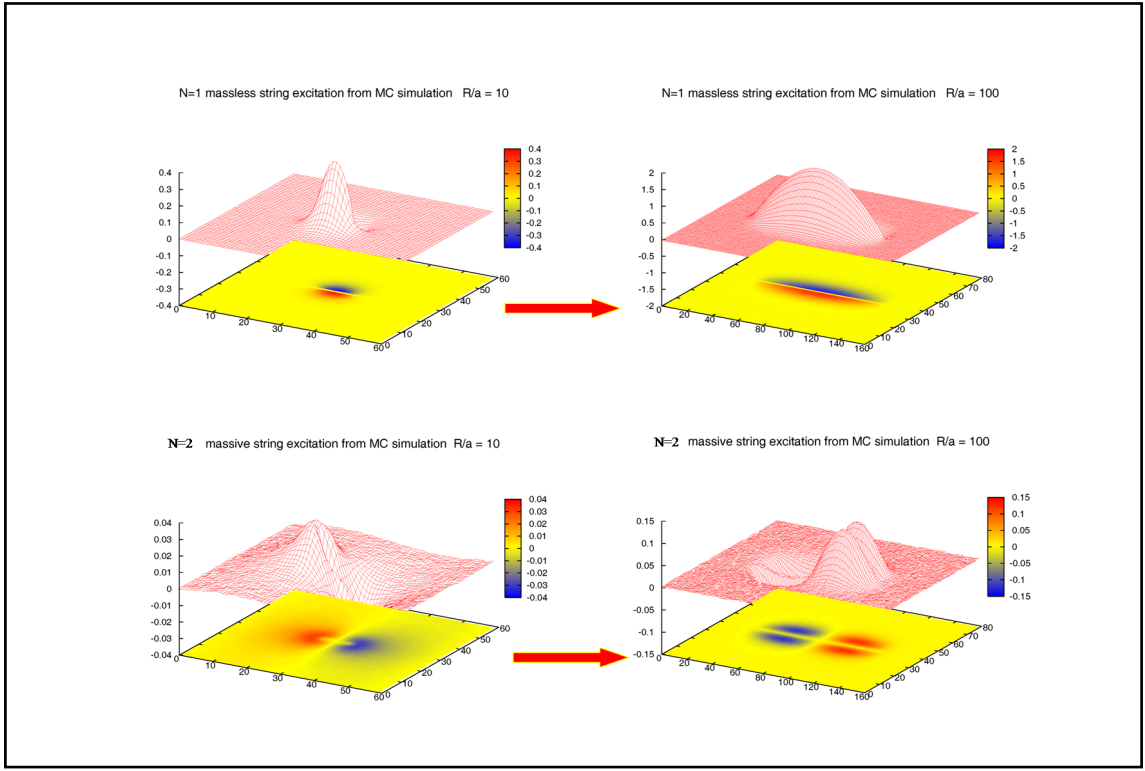

**Figure 2:** Monte Carlo simulation of the first massless Goldstone excitation (upper two wavefunctions) and the second massive breather mode (lower two wavefunctions) in three-dimensional Z(2) lattice gauge theory are shown. The length of the string grows from $R/a = 10$ to $R/a = 100$ in lattice spacing units which on a physical scale, set by the string tension, corresponds to $R$ changing from 0.675 fm to 6.75 fm. Real time animations are constructed from the wavefunctions and energy eigenvalues of Euclidean simulations. Using Adobe Acrobat 6.0, a click within the black frame will start the animation including the N=3 Goldstone excitation.

Since the energy eigenvalues and the wavefunctions of gauge field excitations can be captured in lattice simulations with high accuracy [12], the time-dependent expectation values $\langle \phi(x,y,t) \rangle$ can be followed and animated in real time. The theoretical understanding of the animation is a good starting point for this review.

**2.2 Critical behavior of the Wilson surface**

To explain the origin of string formation in field theory, I will continue the discussion of Z(2) lattice gauge theory with its main features summarized in the chart of Fig. 3. Some important results will also carry over to lattice QCD. Dual transformation maps the Z(2) gauge model into the 3D Ising model with the Z(2) gauge coupling $\overline{\beta}$ related to the dual Ising coupling $\beta$ in the chart [10]. The theory has a nontrivial fixed point $\beta_c$ which separates the confined and deconfined phases of the Z(2) gauge field. The critical region of the bulk confined phase is in the universality class of the three-dimensional $\phi^4$ field theory which describes the fluctuations of the confining electric flux of the closed string-soliton (torelon) with unit winding number in a compact dimension, and the Dirichlet string with static color sources as well. The straightforward generalization of semiclassical soliton quantization in two spatial dimensions is used in the analysis.

The Wilson surface of the Z(2) gauge model exhibits some remarkable behavior as the cou-pling constant is varied. At strong coupling the surface is smooth with a mass gap in its excitation spectrum. At the roughening point $\beta_R$, the surface undergoes a Kosterlitz-Thouless (KT) phase transition [20] where the mass gap disappears and the surface becomes rough with a massless excitation spectrum. The unbinding of screw dislocation pairs in the surface explains the physics of the KT roughening transition. The cut-off dependent interface tension is identified with the string tension which has KT singularity in the gauge coupling at the roughening transition point [21]. Although the Wilson surface remains gapless in its excitation spectrum between the surface roughening transition at $\beta_R$ and the 3d continuum limit fixed point at $\beta_c$, physical properties are expected to change in the $\beta_R \to \beta_c$ crossover.

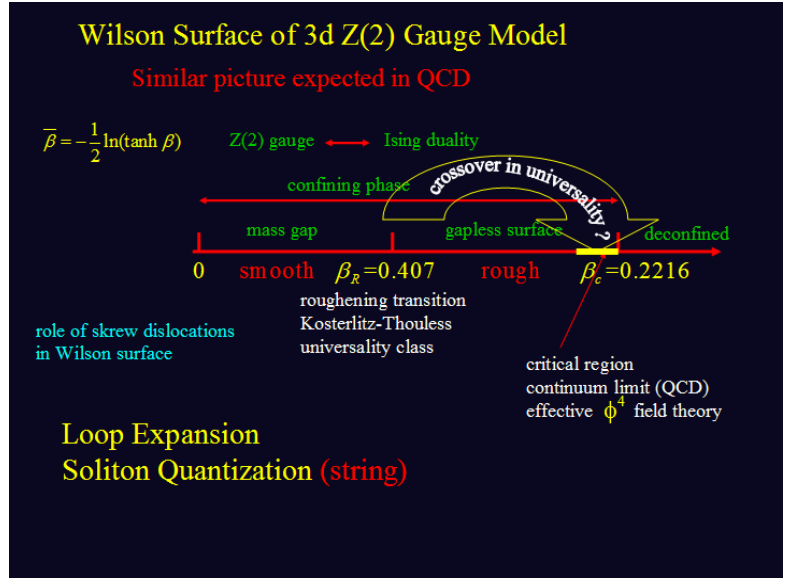

**Figure 3:** Chart summarizing important properties of the three-dimensional Z(2) lattice gauge model and its dual transformation to the Ising model.

*Near the roughening point $\beta_R$*

Near the roughening point, the Wilson surface is embedded in the 3d lattice with small bulk correlation length where rotational symmetry is replaced by cubic symmetry. In the large $L \to \infty$ limit of the spatial extension of the string, with $L$ measured in lattice spacing units, the critical behavior of the Wilson surface is described by the KT universality class. At $\beta_R$ the spectrum is given by the free Gaussian model of two-dimensional conformal field theory with logarithmic corrections from the sine-Gordon operator which describes the periodic rolling of the string on the scale of lattice spacing $a$. The asymptotic Dirichlet string spectrum is given in lattice spacing units by $a \cdot E_n = \pi \cdot n/L$, $n = 1, 2, 3, ...$ with corrections in inverse powers of $\ln L$ which provide a unique signature of the roughening behavior. There are similar logarithmic corrections in the energy spectrum of the winding string with asymptotic levels $a \cdot E_n = 2\pi \cdot n/L$, $n = 1, 2, 3, ...$, at $\beta_R$. Slightly away from the roughening point there are further inverse power corrections to the string spectra coming from higher dimensional lattice operators on the cutoff scale. The logarithmic corrections in the energy spectrum should not be confused with the logarithmic broadening of the surface width at the roughening point which is based on the pure asymptotic form of the spectrum.

*Near the bulk continuum limit $\beta_c$*

Near $\beta_c$, in the confining phase, we do not expect the Wilson surface to exhibit KT behavior. Asymptotically, for large $R$, the torelon energy spectrum is given by $E_n = 2\pi \cdot n/R$, $n = 1,2,3,...$, plus inverse power corrections starting at $O(R^{-3})$. Slow logarithmic corrections are *not* expected. The leading $O(R^{-3})$ correction term appears to be universal, and higher corrections depend on physical properties of the string, like its stiffness and the related massive breather modes. The universality class of the critical point $\beta_c$ controls the scaling properties of surface behavior, string tension, and bulk mass gap (lowest glueball) in the $a \to 0$ lattice spacing limit.

*Crossover in lattice QCD?*

Similar to the Wilson surface in the three-dimensional Z(2) lattice gauge model which exhibits crossover behavior from Kosterlitz-Thouless surface roughening to continuum string theory as the coupling constant is tuned from $\beta_R$ to $\beta_c$, the Wilson surface in SU(N) lattice Yang-Mills models is expected to follow qualitatively similar behavior. The characteristic crossover behavior of the Wilson surface may provide valuable calibration in lattice simulations.

### 2.3 String-soliton quantization

The classical soliton solution $\phi_s$ of the $\phi^4$ field equation in the broken (confining) phase, as depicted in Fig. 4a, represents the ground state of a string with electric flux and one unit of winding number in the compact x-direction. This solution is obtained by imposing antiperiodic boundary

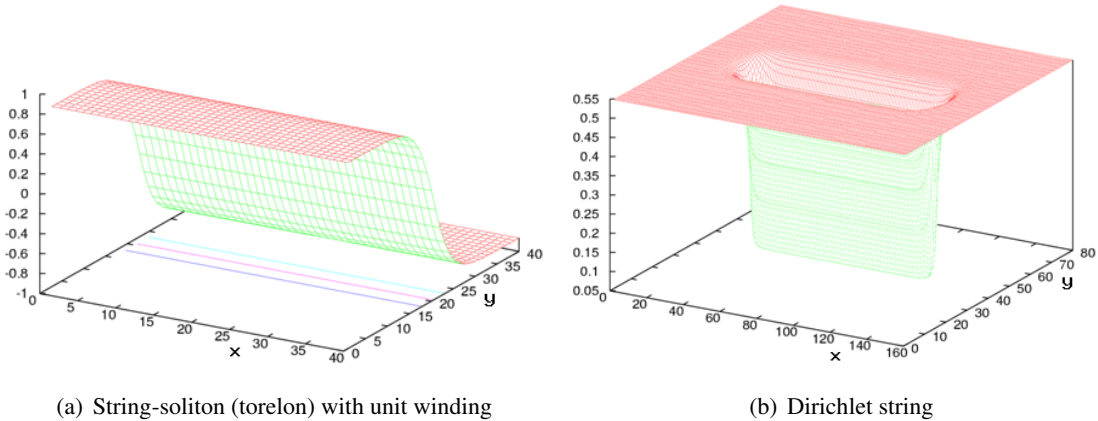

(a) String-soliton (torelon) with unit winding      (b) Dirichlet string

**Figure 4:** Classical string solutions in the dual field variable $\phi$.

condition on the $\phi$ field along the y-direction, or equivalently by a seam of flipped y-links of anti-ferromagnetic Ising couplings lined up parallel in the x-direction. The Dirichlet string with fixed ends is obtained by running a seam of flipped y-links in a finite line segment between static sources along the x-direction, as shown in Fig. 4b. The seams are movable by gauge transformations except the fixed ends. Although the soliton solutions in Fig. 4 are classical in the Ising variables, they represent complicated Z(2) gauge field distributions which would be difficult to understand without the dual representation. In some obvious way, the $\phi^4$ Lagrangian plays the role of the effective Landau-Ginzburg description of quark confinement in dual variables. Many workers would like to use a similar dual representation in Yang-Mills lattice gauge models where it lacks rigorous implementation.

The fluctuation matrix $\mathcal{M} = -\nabla^2 + \mathcal{U}''(\phi_s)$ around the soliton solution $\phi_s$ is derived from the second derivative of the field potential $\mathcal{U}$. The energy eigenvalues and the wavefunctions of the excitations are obtained from the diagonalization of the matrix $\mathcal{M}$. The lattice energy spectrum and wavefunctions around the string-soliton of unit winding factorize,

$$E_{m,n}^2 = \lambda_m^2 + 4\sin^2(q_n/2), \quad \psi_n(x,y) = f_m(y)\exp(iq_n x), \quad q_n = \frac{2\pi n}{L}, \quad n = 1,2,3,..., \qquad (2.1)$$

where the $\lambda_m^2$ eigenvalues and the corresponding $f_m(y)$ eigenfunctions, for $m = 0,1,2,...$, are obtained from the semi-classical quantization of the one-dimensional soliton. There is a tower of states for every value of $m$, with quantized momenta $q_n$, $n = 1,2,3,...$, along the compact x-direction.

The zero mode, $m = 0$, $\lambda_0^2 = 0$, $f_0(y)$, of the one-dimensional soliton describes the translations of string bits in the transverse y-direction and generates massless string excitations with quantized momenta $q_n$, $n = 1,2,3,...$, in the compact x-direction, as shown in Eq. 2.1. The zero mode is responsible for the restoration of broken translational symmetry of the soliton in the y-direction, hence the terminology of massless Goldstone modes of left and right movers. It is important to note that $\lambda_0^2$ is not exactly zero on the lattice due to broken translation invariance. The exponentially small Peierls-Nabarro gap is, however, washed out by quantum fluctuations and the quantum string can smoothly glide through the periodic lattice in continuum energy bands [22].

The next mode, $m = 1$, $\lambda_1^2 = 3M^2/4$, $f_1(y)$, and the spectrum built on it, corresponds to the intrinsic massive breather of the $\phi^4$ soliton which generates the breather modes of the string with a mass gap $\sqrt{3}M/2$ where $M$ designates the lowest glueball mass of the Z(2) gauge theory. $M$ is also identified as the bulk inverse correlation length of the dual field $\phi$. The other dual field excitations, built on the $m = 2,3,...$, towers, are not localized on the confining electric flux and they describe bulk glueball scattering states on the string-soliton.

In the case of the Dirichlet string configuration in Fig. 4b, both the classical 2d soliton solution $\phi_s$ in the dual field variable and the spectrum of $\mathcal{M}$ are obtained from numerical calculations on finite lattices for a fixed separation $R$ of the static sources. In this case there are considerable end distortions in the spectrum and wavefunctions without exact factorization in Eq. 2.1. The effects of the distortions disappear in the $R \to \infty$ limit, but they complicate the matching of the excitation spectra to an effective string Lagrangian in lattice simulations. For a given $R$, the spectrum and the wavefunctions have two conserved quantum numbers with $P_x = \pm 1$, $P_y = \pm 1$ parities for reflection on the x-axis, and y-axis, respectively. String momenta of standing waves are quantized with $q_n = n\pi/R$, $n = 1,2,3,...$, for large $R$.

### 2.4 Origin of collective string variables

The eigenmodes of field fluctuations around the classical solution $\phi_s$ interact in higher orders of the loop expansion. For the string-soliton with unit winding, the form of the analytic solution and the excitation spectrum suggest new field variables $\xi(x,t)$ which are designed to isolate the massless Goldstone fluctuations, together with their interactions, from the rest of the massive field excitations. The new collective field will have the desired geometric interpretation of transverse displacements $\xi(x,t)$ in the y-direction at position x along the string. The coordinate pair $(x,t)$ will become the word sheet parametrization $(\sigma, \tau)$ completing the transfer of massless field fluctuations into physical variables of the geometric string description. We introduce new field variables by writing the field $\phi(x,y,t)$ as

$$\phi(x,y,t) = \phi_s(y - \xi(x,t)) + \eta(x, y - \xi(x,t), t) , \qquad (2.2)$$

where the massive field $\eta$ describes the fluctuations orthogonal to the Goldstone modes, including the massive breathing modes of the string. The unambiguous definition of new collective variables requires the insertion of the Fadeev-Popov identity into the path integral of $\phi^4$ field theory using the constraint $F(\xi(x,t)) = 0$ for the collective variable $\xi(x,t)$,

$$F(\xi(x,t)) = \int_{-\infty}^{\infty} dy \phi_s'(y - \xi(x,t)) \phi(x,y,t) , \quad 1 = \prod_{(x,t)} \int \delta(F(\xi(x,t))) |\frac{\partial F}{\partial \xi}|, \qquad (2.3)$$

where $\phi_s'(y) = \partial \phi_s(y)/\partial y$. The path integral in the new variables $\xi$ and $\eta$ includes the Jacobian of the variable transformation and a rather complicated interaction Lagrangian. Typically, there are derivative interactions like $\partial_a \xi \partial_a \xi \partial_y \eta \partial_y \phi_s$ where $a$ is a summation index over the $(x,t)$ variables which act as world sheet coordinates. To get the effective string theory action, we integrate out the massive field $\eta$. The resulting non-local action of the collective string variable $\xi(\sigma,\tau)$ can be approximated by a derivative expansion under certain well-defined conditions, as illustrated in Fig. 5 where the massive $\eta$ propagator can be replaced by $(\partial^2 \xi)^2$ plus higher derivatives in the limit when the momentum $q$ of the $\xi$ field is much smaller than the mass scale $M$. For the winding string of length $R$, the condition is $q_n = 2\pi \cdot n/R \ll M$ which restricts the validity of the derivative expansion to energy levels with $n \ll M \cdot R/2\pi$ for the torelon. The leading terms of the derivative expansion of the effective Z(2) string action are given by

$$S_{\text{z-string}} = \frac{1}{2\pi\alpha'} \int d\tau \int d\sigma \left\{ 1 - \frac{1}{2}(\partial \xi)^2 - \frac{1}{8}((\partial \xi)^2)^2 + \mathcal{O}(((\partial \xi)^2)^3) \right\} + S_{FP}(\xi) , \qquad (2.4)$$

where $1/2\pi\alpha'$ is the string tension which is determined by the energy of the string-soliton per unit length. Contributions from the Fadeev-Popov determinant, represented by $S_{FP}(\xi)$, only effect derivatives starting in sixth order.

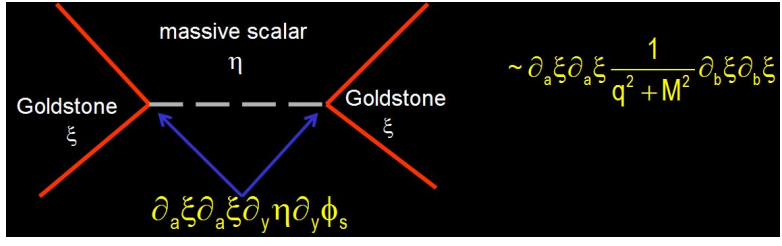

**Figure 5:** The massive $\eta$ propagator with derivative interaction vertices leads to the effective local derivative expansion of the string Lagrangian for $\xi(\sigma,\tau)$.

The first three terms of the effective action in Eq. 2.4 match the first three terms of the $(\partial \xi)^2$ expansion of the 3D classical Nambu-Goto string action given by $S_{NG} = \sqrt{1 + (\partial \xi)^2}$ in our variables. In higher orders, the expansion of $S_{\text{z-string}}$ deviates from $S_{NG}$. For example, some sixth order derivative terms in $S_{\text{z-string}}$ are expected to match the extrinsic geometric curvature term introduce by Polyakov [23] and related to the negative stiffness of the Z(2) string [24]. It appears that the extrinsic curvature term in the action, the negative stiffness of the string, and the presence of massive breather modes are closely related physical properties of the confining flux in the three-dimesional Z(2) gauge theory. I expect that similar features remain tightly connected in other string models as well, but the Nambu-Goto string is known to have zero stiffness.

### 2.5 Lüscher-Weisz effective string action

The effective string action derived in Eq. 2.4 should be universal up to fourth order derivatives for string formation in 3d field theories, including Yang-Mills gauge models. The only parameter to fourth order is the string tension, and new physics, which may not be universal, would be coded in higher order derivatives. I expect that the fourth order derivatives remain universal in D=4 dimensions where two independent fourth order derivative couplings can be constructed. We also want to know how to handle end distortions when Dirichlet boundary conditions are imposed. In fact, the role of boundary operators has to be addressed in three dimensions as well.

Designed to be independent from field theory specifics, Lüscher and Weisz (LW) developed a string effective action which is applicable in any dimension D, with boundary operators designed to describe end distortions of the Dirichlet string [25]. The LW effective action, consistent with string symmetries, has the form

$$S_{LW} = \sigma RT + \mu T + \frac{1}{2\pi\alpha'} \int_0^T d\tau \int_0^R d\sigma \frac{1}{2} \partial_a \xi \partial_a \xi + \frac{1}{4} b \int_0^T d\tau \left\{ (\partial_1 \xi \partial_1 \xi)_{\sigma=0} + (\partial_1 \xi \partial_1 \xi)_{\sigma=R} \right\}$$

$$+ \frac{1}{4} c_2 \int_0^T d\tau \int_0^R d\sigma \frac{1}{2} (\partial_a \xi \partial_a \xi)(\partial_b \xi \partial_b \xi) + \frac{1}{4} c_3 \int_0^T d\tau \int_0^R d\sigma \frac{1}{2} (\partial_a \xi \partial_b \xi)(\partial_a \xi \partial_b \xi) + ..., \quad (2.5)$$

to fourth order in the derivatives (the string displacement vector $\xi$ has D-2 components and the self-energy mass term $\mu T$ is irrelevant for our discussion). A duality argument was invoked to set the boundary coupling to $b=0$, in addition to a duality relation between $c_2$ and $c_3$ [25]. In D=4 dimensions, this leaves one independent interaction term with arbitrary coupling in fourth order. In D=3 dimensions, the two derivative interactions in fourth order are kinematically not independent, both $c_2$ and $c_3$ are fixed by symmetries, and the effective string action $S_{LW}$ in Eq. 2.5 becomes identical to $S_{z\text{-string}}$ in Eq. 2.4 up to and including fourth order derivatives.

The effective string actions $S_{LS}$ and $S_{z\text{-string}}$ are very similar in their scope of use. They are built in the transverse gauge with D-2 string oscillators and do not have full Poincare invariance in the strict string theory sense implying that boosts to other Lorentz frames are not built into their descriptions. The fully relativistic spectrum of a closed string-soliton, or torelon, cannot be reproduced without proper treatment of the center of mass string motion. It also remains unclear whether the lack of reparameterization invariance of the string action in Eq. 2.5 has additional consequences. In the rest of this review I will discuss the lattice simulation results using the non-critical effective string action developed by Polchinski and Strominger [7]. This approach has full Poincare invariance and reparameterization symmetry on the world sheet circumventing the quantization problems of the Nambu-Goto string in non-critical physical dimensions.

## 3. Poincare invariant effective string theory with reparameterization symmetry

To develop an effective action of the collective string coordinates with Poincare invariance and reparameterization symmetry, we consider the one-dimensional string as it sweeps out a two-dimensional world-sheet surface described by $X^\mu(\tau,\sigma)$. The $(\tau,\sigma)$ world sheet coordinates map into string coordinates $[X^0(\tau,\sigma), X^1(\tau,\sigma),...,X^d(\tau,\sigma)]$ in $D = d+1$ dimensional Minkowski space-time with negative time-like sign in the metric tensor, in the metric sign convention I adapt in this section. Consistent relativistic quantum theory requires parameterization invariance while the string action should only depend on embedding in space-time which is characterized by the induced metric $h_{ab} = \partial_a X^\mu \partial_b X_\mu$ where a,b refer to the $(\tau,\sigma)$ world sheet coordinates. The simplest choice for the string Lagrangian is the Nambu-Goto action,

$$S_{NG} = \int_M d\tau d\sigma \, \mathscr{L}_{NG}, \quad \mathscr{L}_{NG} = -\frac{1}{2\pi\alpha'}(-det\, h_{ab})^{1/2}, \tag{3.1}$$

where $1/2\pi\alpha'$ is identified with the string tension and $M$ designates the world sheet swept out by the string. On the classical level, the action is invariant under the D-dimensional Poincare group and it also has two-dimensional coordinate (Diff) invariance. However, there is a well-known problem with the quantization procedure. Quantization in light cone gauge has the correct number of D-2 transverse string oscillators but Poincare invariance is lost outside the critical D=26 dimensions. Covariant (Virasoro) quantization exhibits D-1 oscillators unless D=26.

Trying to remedy the problem, Polyakov introduced word sheet metric $\gamma_{ab}(\tau,\sigma)$ into the string action,

$$S_{Pol}[X,\gamma] = \int_M d\tau d\sigma \, (-\gamma)^{1/2} \gamma^{ab} \partial_a X^\mu \partial_b X_\mu, \quad \gamma = det\, \gamma^{ab}. \tag{3.2}$$

Weyl equivalent world sheet metrics with $\gamma'^{ab} = exp(2\omega(\tau,\sigma))\gamma^{ab}$ and an arbitrary $\omega(\tau,\sigma)$ function correspond to the same embedding in space-time. In addition to D-dimensional Poincare invariance and two-dimensional coordinate (Diff) invariance, there is now also a new two-dimensional Weyl symmetry and Polyakov's quantization runs into a difficulty. There is a new Liouville mode (scalar field) in the path integral which leads again to D-1 oscillators instead of D-2.

On the classical level, $S_{Pol}$ is equivalent to $S_{NG}$. On the quantum level, Polchinski and Strominger (PS) suggested a new way to construct an effective string theory of long flux lines emerging from quantum field theories. In the Jacobian of the path integral of collective variables, they replaced in Polyakov's approach the intrinsic metric with the embedding metric, which should keep the effective string as close as possible to its field theory origin and the NG form. Their construction requires two major steps. First, it is noted that starting from the path integral of field theory, the collective coordinate quantization of the long flux line will generate measure terms which will appear as the determinant of the Jacobian related to the change of variables. The collective string coordinates of D unconstrained $X^\mu$ fields lead to the string action

$$S_0 = \frac{1}{2\pi\alpha'} \int d\tau^+ d\tau^- \partial_+ X^\mu \partial_- X_\mu + \text{determinant}. \tag{3.3}$$

In the second step of the construction, it is noted that the measure derives from the physical motion of the underlying gauge fields and therefore should be built from physical objects, like the induced metric $h_{ab} = \partial_a X^\mu \partial_b X_\mu$. At this point the conjecture is made that the determinant should have the same form as in Polyakov's path integral quantization but built from the induced metric. The Polyakov determinant in conformal gauge has the form $e^{iS_L}$ where in terms of Polyakov's intrinsic metric $e^\phi$, the substitution of the induced metric leads to

$$S_L = \frac{26-D}{48\pi} \int d\tau^+ d\tau^- \partial_+ \phi \partial_- \phi \Longrightarrow \frac{26-D}{48\pi} \int d\tau^+ d\tau^- \frac{\partial_+^2 X \partial_- X \partial_+ X \partial_-^2 X}{(\partial_+ X \partial_- X)^2}. \tag{3.4}$$

After substituting the induced conformal gauge metric $h_{+-}$ for $e^\phi$, one finally obtains the full effective action

$$S_{PS} = \frac{1}{4\pi} \int d\tau^+ d\tau^- \left[ \frac{1}{\alpha'/2} \partial_+ X^\mu \partial_- X_\mu + \beta \frac{\partial_+^2 X \partial_- X \partial_+ X \partial_-^2 X}{(\partial_+ X \partial_- X)^2} \right] + ..., \quad (3.5)$$

where $\beta = \beta_c = \frac{D-26}{12}$ is required for the anomaly free theory. The action has manifest D-dimesional Poincare invariance and conformal symmetry. It is an effective theory only, because additional terms may be required when expanded in derivative powers of the long string-soliton. The obvious example again is the stiffness, or extrinsic curvature term, which will require added terms in Eq. 3.5. Further higher derivative terms in Eq. 3.5 are also required to maintain conformal invariance beyond fourth order derivatives.

Based on Eq. 3.5 of the PS effective action, the asymptotic form of the string-soliton spectrum with winding number $w$ along the compact dimension was worked out in three independent calculations [26, 27]. The spectrum is given by the formula

$$E^2_{n,N+\widetilde{N}} = \sigma^2 R^2 w^2 - \frac{\pi}{3} \sigma(D-2) + 4\pi\sigma(N+\widetilde{N}) + \frac{4\pi^2 n^2}{R^2} + \vec{p}_T^2 , \quad (3.6)$$

where $N+\widetilde{N}$ describes the sum of right and left movers along the compactified dimension of length $R$. The momentum square $\vec{p}_{n\parallel}^2 = (2\pi n/R)^2$, $n = 0, 1, 2, ...,$ along the compactified direction is quantized, and for any integer winding number $w$ there exists a relation $N - \widetilde{N} = nw$ between the discrete compact momentum and the difference in the number of right and left movers. The transverse momentum $\vec{p}_T$ of the winding string is a D-2 dimensional vector, with $\vec{p}_T = 0$ in the simulations of $w = 1$ torelon spectra which I will discuss later. Inverse power corrections to $E^2_{n,N+\widetilde{N}}$ starting at order $\mathcal{O}(R^{-3})$ in Eq. 3.6 will be determined by the systematic restoration of conformal symmetry in higher orders of the derivative expansion of the nonlinear effective action around the long string-soliton, and by new physical properties of the string. It remains unknown how Poincare invariance will be maintained in the presence of new correction terms. The spectrum of the Dirichlet string can also be calculated by imposing boundary conditions on the effective action of Eq. 3.6. The result is given by

$$E_N = \sigma R \sqrt{1 - \frac{\pi}{12\sigma R^2}(D-2) + \frac{2\pi N}{\sigma R^2}} , \quad (3.7)$$

which agrees with the old result of Arvis [28] apart from unknown correction terms to Eq. 3.7 starting at order $O(R^{-5})$.

The results in Eq. 3.6 and Eq. 3.7 have interesting properties. Without added further correction terms they agree with the Nambu-Goto predictions for the excitation energies. At smaller values of $R$ there will be deviations from the NG spectrum, before the scale of the tachyon singularity is reached but the required correction terms remain unknown. The torelon spectrum exhibits full relativistic invariance with $E^2_{n,N+\widetilde{N}} = M^2_{N+\widetilde{N}} + \vec{p}_n^2$ where $M^2_{N+\widetilde{N}} = \sigma^2 R^2 w^2 - \frac{\pi}{3}\sigma(D-2) + 4\pi\sigma(N+\widetilde{N})$ designates the Poincare invariant squared mass of the excited torelon state and $\vec{p}_n = (2\pi n/R) \cdot \vec{e}_\parallel + \vec{p}_T$, with the unit vector $\vec{e}_\parallel$ pointing in the compact direction. To $O(R^{-3})$ order the only free parameter in Eq. 3.7 is the string tension $\sigma = 1/2\pi\alpha'$. The result is consistent with the calculation of Lüscher and Weisz [25] but the free couplings $c_2, c_3$ are both fixed by the symmetries. The torelon spectrum in Eq. 3.6 has to be used in the same fashion. After taking the square root of the right-hand side and reexpanding in odd powers of $R^{-1}$, a universal prediction $\mathcal{O}(R^{-3})$ appears for large $R$, but $O(R^{-5})$ terms will depend on new string physics. The proper treatment of boundary operators remains unknown in the PS formulation.

## 4. Casimir energy of the string ground state

The Casimir energy of the confining flux is identified as the zero point energy of string oscillators in high precision measurements of the ground state energy, an interesting and popular topic with recent lattice results for the bosonic Dirichlet string and the baryon string with Y-junction.

### 4.1 Casimir energy of the bosonic string

The effective string prediction for the ground state energy of two static sources at separation $R$ is given asymptotically by

$$E_0(R) = \sigma R - \frac{\pi(D-2)}{24R} - \frac{\pi^2(D-2)^2}{1152\sigma R^3} \ , \tag{4.1}$$

where the model dependent correction $\mathcal{O}(R^{-5})$ and a cutoff dependent constant, not contributing to $E_0'(R)$, are not included. There is strong evidence, without rigorous proof, that the three terms of $E_0(R)$ are universal for bosonic strings: (1) they are known to match field theory calculations around string-like soliton solutions, (2) the three terms are also matched in Nambu-Goto string quantization, consistent on the $\mathcal{O}(R^{-3})$ scale [28], (3) equivalent three terms were derived for the string-soliton with unit winding from the Poincare invariant effective string action with reparameterization symmetry [26, 27]. Eq. 4.1 would also follow from the $S_{LW}$ action of Eq. 2.5, if the free coupling were set to $c_2 = \frac{1}{2\sigma}$, as anticipated by other string symmetries not considered in [25]. Corrections to $E_0(R)$, starting at $\mathcal{O}(R^{-5})$, are not predictable without new terms and parameters in the effective string action. New physics, like string stiffness, will be coded in the higher order terms. Deviations are certainly expected from the Nambu-Goto form which has no new free parameters but becomes inconsistent on the $\mathcal{O}(R^{-5})$ scale in non-critical physical dimensions [28].

Taking the second derivative of $E_0(R)$, the effective Casimir term, defined by $C_{eff}(R) = -\frac{1}{2}R^3 E_0''(R)$, has the asymptotic behavior $C_{eff}(R) = \frac{\pi(D-2)}{24}(1 + \frac{\pi(D-2)}{8\sigma R^2} + ...)$ which isolates the Casimir energy with leading $R^{-2}$ correction. The definitive lattice simulation of $C_{eff}(R)$ came from an efficient method to measure the ground state energy with high precision [29] as shown in Fig. 6. The results had the popular interpretation of bosonic string formation on the scale $R \lesssim 1$ fm. In D=3 dimensions the agreement seemed to be a perfect match to $\pi/24$, but boundary operators had to be added from Eq. 2.5 to explain deviations from $\pi/12$ when D=4 [29]. In fact, the interpretation should have been the reverse. The universal $O(R^{-2})$ asymptotic behavior, represented by the dashed green line in Fig. 6, is a measurable correction to $\pi/12$ and apparently reached by the data for D=4. Therefore the D=4 result does not require any boundary terms which were later ruled out by LW duality regardless [25]. It is in D=3 dimensions that the lattice results remain noticeably below the universal dashed green line.

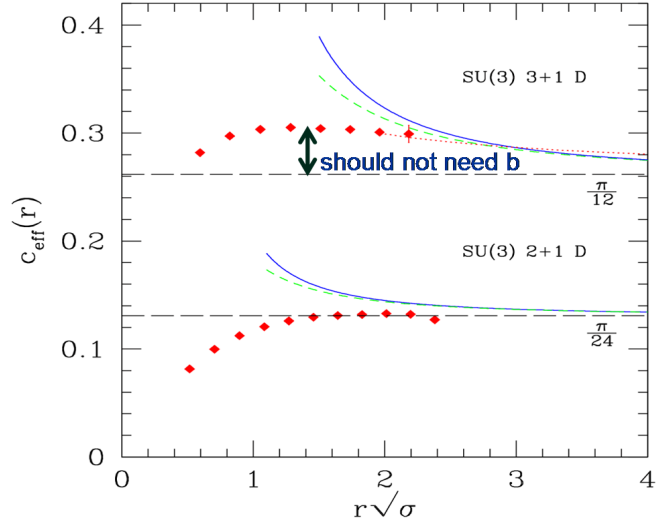

**Figure 6:** The dashed green lines designate $E_0(R)$ of Eq. 4.1. The solid blue lines are the NG predictions from the full square root expression of Eq. 3.7. The red data points and the dotted red line, which includes boundary operator corrections, are taken from [29].

Although the observed differences in trend for D=3 and D=4 remain unexplained, there are three effects which can delay the onset of asymptotic behavior in $C_{eff}(R)$ in increasing order of likelihood: (1) large $O(R^{-5})$ effects, (2) interactions of exact massless Goldstone modes outside the range of the derivative expansion, (3) end distortions by the static sources which effect the spectrum of the Goldstone modes before they fully develop into massless excitations. It remains unclear how to disentangle the last effect from asymptotic string behavior with or without some kind of appropriate boundary operator description, if it exists. In addition, there is always the question of lattice cutoff effects which is not my main concern in this particular case.

Useful new results were obtained on the asymptotic behavior of $C_{eff}(R)$ in 3D Z(2) and SU(2) gauge models, with results shown in Fig. 7. Trends in the new lattice data are remarkably close to the 3D SU(3) model.

*Conclusions from Casimir energy studies of the bosonic string:* (1) Confirmation of bosonic string formation from high precision Casimir energy measurements in D=3 dimensions would require reach into the $R = 2$ fm range, (2) Disentangling end effects may remain a serious conceptual and practical problem, (3) isolating the Casimir energy of the torelon ground state with high precision, comparable to the Dirichlet string, is important for eliminating boundary problems. [1]

### 4.2 Casimir energy of the baryon string

When three static sources are inserted in a baryon configuration, a three-string junction with Y-shape is expected to form asymptotically. The baryon string in the 4D SU(3) lattice Yang-Mills theory was investigated [33] and it was argued that the best interpretation of the simulations corresponds to the anticipated Y-shape three-string configuration, with the first gluon excitation of the three-string also reported [33]. Observation of the baryon string Casimir energy associated with the Y-shape three-string configuration was reported in the 3D Z(3) gauge model [34]. The result, shown in Fig. 8a, is surprising, given the model assumptions gong into the analytic Casimir energy calculation and without knowing the baryon string spectrum [34].

---
[1]Without high precision in $E'_0(R)$, a global fit to $E_0(R)$ was reported consistent with the torelon Casimir energy [32].

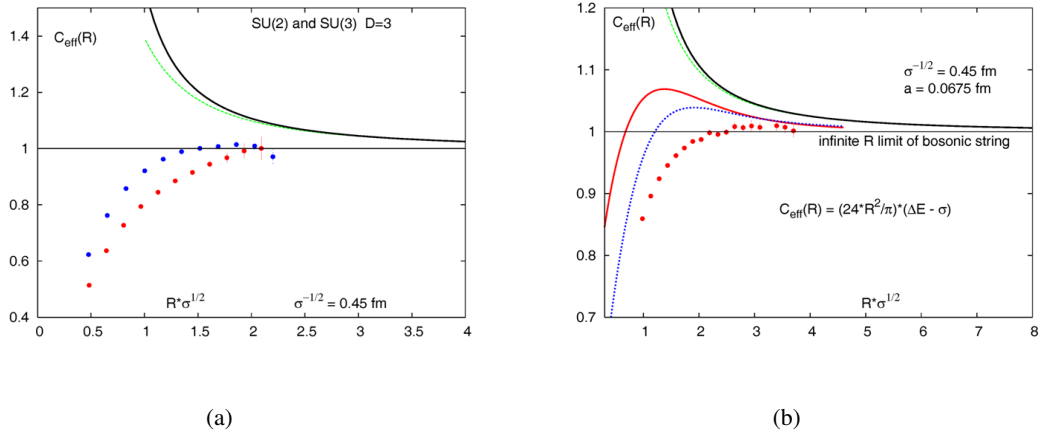

**Figure 7:** (a) The red point show $C_{eff}$, asymptotically normalized to one, for the 3D SU(2) gauge group [30, 31]. The blue points for the 3D SU(3) model are shown for comparison [29]. The full NG prediction (solid black line) and the dashed green line were defined earlier in Fig. 6. (b) Results for the 3D Z(2) gauge model are shown with a modified definition of $C_{eff}(R)$ using finite lattice difference $\Delta E$ for $E'_0(R)$, with the contribution of the string tension removed [12, 13]. The blue dotted line is the zero-point energy contribution of the field oscillators. The solid red line also includes the contribution from the classical energy of the $\phi_s$ soliton solution [12].

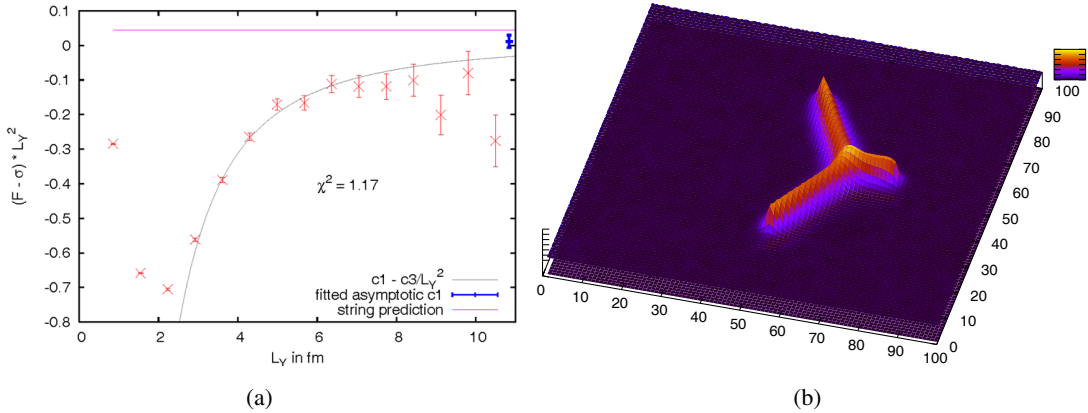

**Figure 8:** (a) shows the isolation of the the ground state Casimir energy of the baryon string when three static sources are located in equilateral configuration. The force was measured in the 3D Z(3) gauge model with the asymptotic string tension of a Y-shape three-string subtracted and comparison was made with an analytic model which predicts vanishing Casimir energy asymptotically. The shape of the $(F-\sigma)\cdot L_Y^2$ data at gauge coupling $\beta = 0.6$ was interpreted as evidence for Y-shape three-string formation [34]. (b) Graphical evidence for baryon string formation in the 3D Z(3) gauge model is shown at gauge coupling $\beta = 0.59$ [35].

    Two questions stand out: (1) Is the asymptotic baryon string really Y-shaped asymptotically? (2) What is the spectrum and Casimir energy of the Y-shape three-string from first principles string theory calculation and lattice simulations? In collaboration with Holland I looked into the space-

time picture of baryon string formation in the 3D Z(3) gauge model [35]. We find convincing evidence for baryon three-string formation with Y-junction in Fig. 8b, continuing the investigation with the baryon string spectrum and the related baryon Casimir energy.

## 5. Dirichlet string spectrum

Recent studies of the Dirichlet string spectrum include the 3D Z(2) gauge model [12, 13, 14], the 3D SU(2) Yang-Mills model [36, 31], and the 4D SU(3) gluon sector of lattice QCD [37, 38].

### 5.1 Dirichlet string spectrum of the 3D Z(2) gauge model

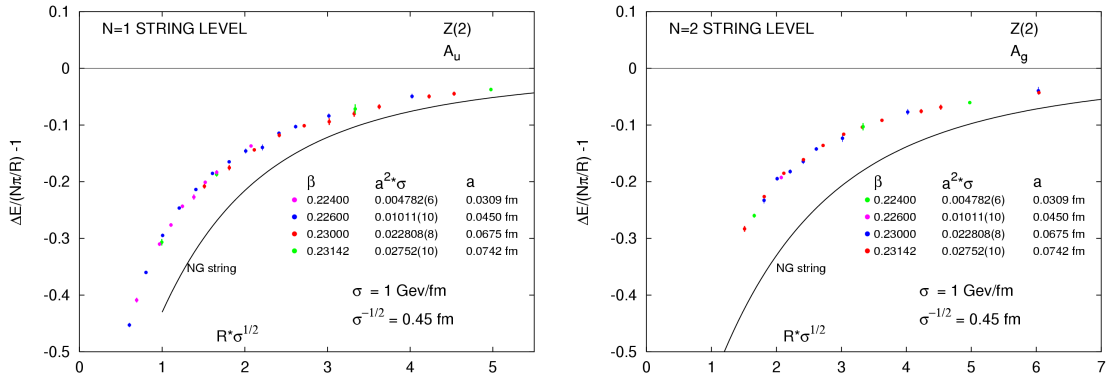

**Figure 9:** Evidence for scaling in the string spectrum is shown for the $N = 1, 2$ Dirichlet string states ($A_u$ and $A_g$ are point group notations) in the indicated range of lattice spacings [13]. The solid black reference line is the full NG prediction which is expected to break down at small $R$ values.

The simulation of the Dirichlet string spectrum was implemented in the dual Ising representation of the 3D Z(2) gauge model with two static sources built into the action by a seam of flipped y-links running along the x-direction of the lattice, as discussed in section 2. The seam of links is repeated on every spatial slice of the 3d lattice creating a surface along the Euclidean time direction with periodic wrapping. This is equivalent to a pair of Polyakov loops built into the importance sampling in the gauge theory representation of the action. In the presence of the fixed Polyakov loops, the plaquette-plaquette correlators of the Z(2) gauge field are measured to capture string excitations. This method is equivalent to full information on the connected correlation function of four operators which include two Polyakov loops and two plaquette operators. A large set of string operators was constructed in the dual Ising representation to define a correlation matrices in each of the four channels characterized by two exact parity quantum numbers $(P_x, P_y)$. The operators were based on wavefunctions of field excitations around the classical string solution $\phi_s$, as described in section 2.

With this method, several excitations were determined with great precision in each of the four channels [12, 13]. Fig. 9 shows the first two Dirichlet string energy levels. The energy gaps $\Delta E_N$ were directly measured above the ground state and the quantity $R \cdot \Delta E_N / N\pi - 1$ is plotted to show percentage deviations from the asymptotic Dirichlet string levels for string quantum numbers $N = 1, 2$. Each $R$ value requires an independent run which generates completely decorrelated R-shapes for the excitation curves. Data from runs at four different gauge couplings are shown in Fig. 9 with good scaling properties. The string tension was determined for each gauge coupling in separate runs using a method designed for accurate measurements of the free energy of the Wilson surface between two Polyakov loops [12, 13]. With good scaling demonstrated in Fig. 9, Dirichlet string theory for the lowest four string excitations is tested in Fig. 10 at one selected value of the gauge coupling with reach to $R = 7$ fm separation. The analysis of lattice simulations in effective string theory is very similar to the discussion of section 4.1 for the ground state energy.
*Conclusions from the 3D Z(2) Dirichlet string spectrum:* (1) For large values of $R$ the expected asymptotic string limit has been reached, (2) the energy spectrum of the fluctuation operator $\mathcal{M}$ of section 2 indicates the presence of noticeable end effects in the spectrum, (3) it remains unclear whether the end effects can be described by some string boundary operators, (4) without knowing the R-dependence of end distortions, it is difficult to disentangle boundary effects from asymptotic $O(R^{-3})$ behavior in the energy gaps which makes it problematic how to probe new physics in the effective string action.

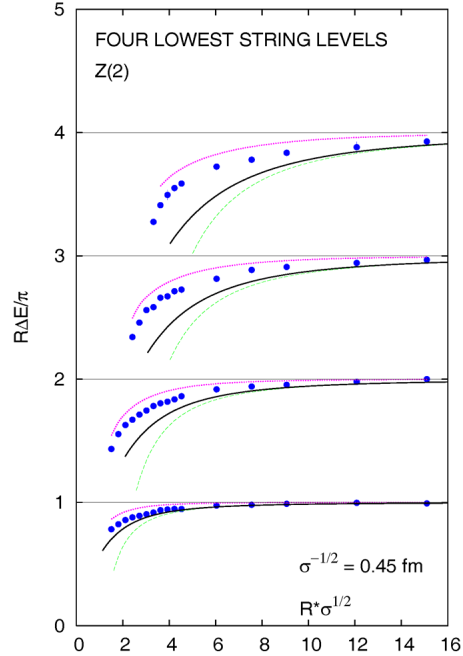

**Figure 10:** $R \cdot \Delta E / \pi$ is plotted for the first four Dirichlet string excitations in the 3D Z(2) gauge model at coupling $\beta = 0.23$ in the dual Ising representation. Solid black lines show the full NG prediction which will break down at small $R$ values. Dashed green lines represent universal predictions, accurate on the scale $\mathcal{O}(R^{-3})$ of the effective string action, as explained earlier. On the scale $O(R^{-5})$ string predictions become model dependent. Dotted lines show the energy eigenvalues as calculated numerically from the fluctuation operator $\mathcal{M}$ of section 2 in soliton quantization. Their deviations from the integer lines are good indicators of end distortion effects [13].

### 5.2 Dirichlet string spectrum of the 3D SU(2) gauge model

It is important to check now that the trends found in the Z(2) gauge model remain similar in the 3D SU(2) Yang-Mills model which is a step closer to lattice QCD with Z(2) center group. The excitation energies of two Dirichlet string states from large correlation matrices of Wilson operators with string quantum numbers are shown in Fig. 11 in 3D SU(2) Yang-Mills gauge theory [36]. Results for the lowest N=1 state were also obtained in a different simulation [31]. Unfortunately, the large errors in [31] makes useful comparisons with the high precision work of [36] problematic. *Conclusions from the 3D SU(2) Dirichlet string spectrum:* (1) The important trends established in the Z(2) gauge model remain very similar in the SU(2) model which maybe related to the center group, (2) for large values of *R* the expected asymptotic string limit has been reached but the scales probed are more limited than in Z(2), (3) interpretation of boundary effects remain a problem as it was discussed earlier.

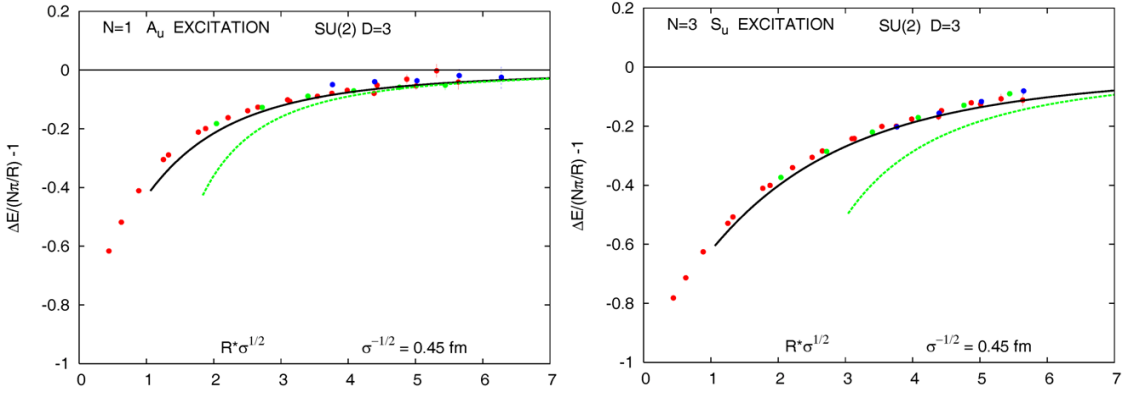

**Figure 11:** The N=1 and N=3 Dirichlet string excitations are displayed in the 3D SU(2) model with reach to R=3 fm. The definitions of the solid black lines and the dashed green lines were given in Fig. 10.

### 5.3 Dirichlet string spectrum of the 4D SU(3) gauge model

In the 4D SU(3) Yang-Mills gauge model, which is the gluon sector of QCD, three exact quantum numbers determine the classification scheme of the gluon excitation spectrum in the presence of a static $q\bar{q}$ pair [37]. We adopt the standard notation from the physics of diatomic molecules and use $\Lambda$ to denote the magnitude of the eigenvalue of the projection of the total angular momentum of the gluon field onto the molecular axis. The capital Greek letters $\Sigma, \Pi, \Delta, \Phi, \ldots$ are used to indicate states with $\Lambda = 0, 1, 2, 3, \ldots$, respectively. The combined operations of charge conjugation and spatial inversion about the midpoint between the quark and the antiquark is also a symmetry and its eigenvalue is denoted by $\eta_{CP}$. States with $\eta_{CP} = 1(-1)$ are denoted by the subscripts $g$ ($u$). There is an additional label for the $\Sigma$ states; $\Sigma$ states which are even (odd) under a reflection in a plane containing the molecular axis are denoted by a superscript $+$ ($-$). Hence, the low-lying levels are labeled $\Sigma_g^+, \Sigma_g^-, \Sigma_u^+, \Sigma_u^-, \Pi_g, \Pi_u, \Delta_g, \Delta_u$, and so on [37]. Restricted to the $R = 0.2 - 2$ fm range of selected simulations, energy gaps $\Delta E$ above the ground state are compared to asymptotic string gaps for two selected excited states in Fig. 12. The quantity $R \cdot \Delta E / N\pi - 1$ is plotted to show percentage deviations from the asymptotic string levels.

In sharp contrast, for $R \ll 1$ fm it was shown that the multipole expansion of the gluon field can be applied with success in the short-distance operator product expansion of gluon excitations around static color sources [37, 38]. The observed short distance level ordering is very different from the string spectrum, as illustrated in Fig. 13. For 0.5 fm $< R <$ 2 fm, a dramatic crossover of the energy levels toward a string-like spectrum is observed as $R$ increases. For example, the N=3 $\Sigma_u^-$ state breaks rapidly away from its N=1 $\Pi_u$ short-distance $O(3)$ degeneracy partner to approach the ordering and degeneracies expected from bosonic string theory.

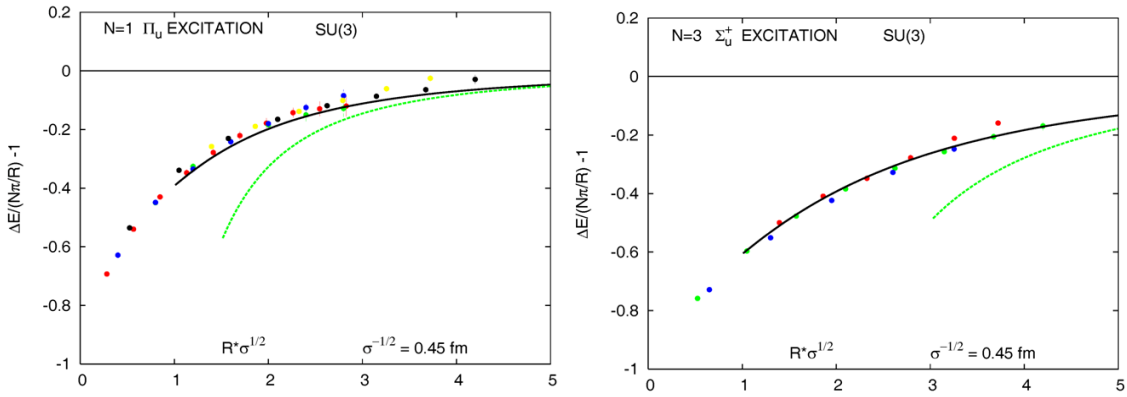

**Figure 12:** Two Dirichlet string excitations are displayed in the 4D SU(3) gauge model with reach to R=2 fm. Solid black lines show the full NG prediction which will break down at small *R* values. The colors of the data points identify the gauge coupling of the corresponding run, demonstrating scaling behavior in the spectrum. Dashed green lines represent universal predictions on the $\mathcal{O}(R^{-3})$ scale of the effective string action, as it was discussed earlier.

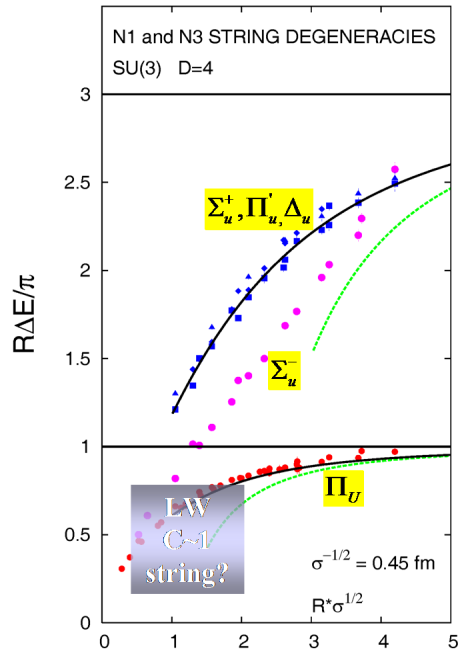

**Figure 13:** The red data points correspond to the $\Pi_u$ gluon excitation which is the only N=1 string state in the plot. There are four states in the plot which are N=3 string states asymptotically: $\Sigma_u^+$(blue squares), $\Pi'_u$(blue diamonds) which is the first radial excitation with $\Pi_u$ quantum numbers, $\Delta_u$(blue triangle), $\Sigma_u^-$ (cyan circle). The multipole expansion predicts the non-string degeneracy of $\Sigma_u^-$ and $\Pi_u$ at short distances. For large R, the $\Sigma_u^-$ state breaks away and joins its degenerate N=3 string partners. The definition of the solid black line and the dashed green line are the same as in Fig 12. The Casimir data of [25] are located in the region where $\Sigma_u^-$ is just beginning to break away from its short distance non-string behavior.

*Conclusions from the 4D SU(3) Dirichlet string spectrum:* (1) There is a dramatic crossover from short distance physics to string physics in the spectrum, (2) it will remain very difficult to disentangle $O(R^{-3})$ string effects from boundary effects in the Dirichlet spectrum when new physics is probed with the effective string action, (3) there is strong motivation to turn to the torelon spectrum which is free of end distortions.

## 6. Torelon string spectrum

Two recent studies investigated the spectrum of the torelon. A high precision study of the 3D Z(2) gauge model is reported here for the first time [13], together with the torelon spectrum of 4D SU(3) Yang-Mills gauge theory which is the gluon sector of QCD [27, 39].

### 6.1 Torelon spectrum of the 3D Z(2) gauge model

The simulation of the torelon spectrum is a straightforward application of the procedure outlined in section 5.1. The seam creating the torelon runs across the periodic x-direction of the lattice creating one unit of winding with $w = 1$. This seam can be moved by gauge transformations and the torelon is free to be moved in the y-direction, with zero transverse momentum projected. The ex-citation spectrum of the torelon should be compared with the predictions of effective string theory as given in Eq. 3.6. In Fig. 14a the $n = 1, N = 1, \widetilde{N} = 0$ energy level is plotted. Fig. 14b shows two torelon states with the same principal quantum number, $N + \widetilde{N} = 2$, but different $\vec{p}_n$ momenta. A single right mover with $N = 2, \widetilde{N} = 0$ and running with two units of momentum in the compactified direction has positive $\mathcal{O}(R^{-3})$ energy correction to the asymptotic string level and split from the state of the composite state of back to back right and left movers with zero total momentum in the compact dimension. The dramatic fine structure, predicted in Eq. 3.6, is in impressive agreement with lattice simulations [13].

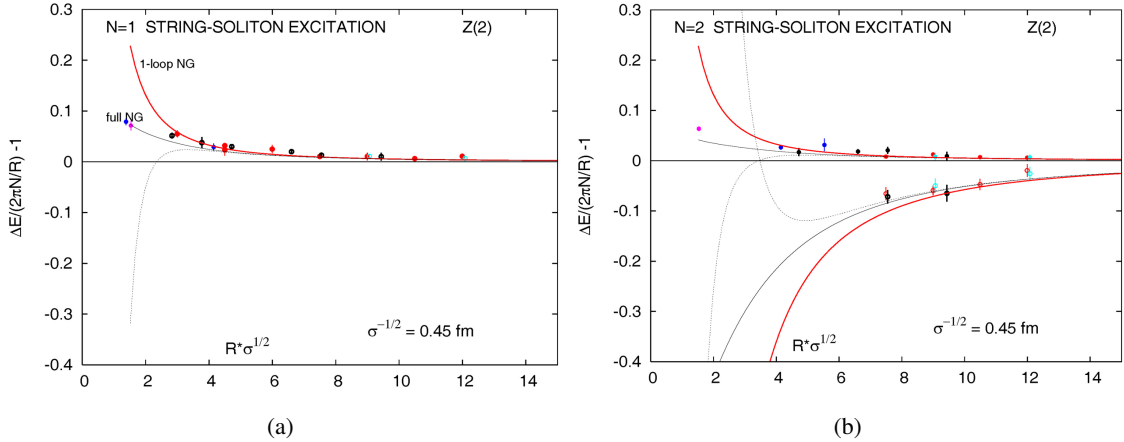

**Figure 14:** The spectrum in (a) corresponds to the $w = 1, N = 1, \widetilde{N} = 0, p = 2\pi/R$ state which is the lowest torelon excitation with $N + \widetilde{N} = 1$. The spectra in (b) correspond to two different torelon excitations with $w = 1, N + \widetilde{N} = 2$. Above the null line the spectrum of the $n = 2, N = 2, \widetilde{N} = 0$ state is depicted with $p_n = 4\pi/R$ momentum in the compact dimension. Below the null line, the spectrum of the state $N = 1, \widetilde{N} = 1, p_n = 0$ is shown. The red line in both figures corresponds to the spectrum of Eq. 3.6 which coincides with the exact NG prediction. The black lines show the universal $\mathcal{O}(R^{-3})$ predictions of the effective string action after the square root of the right hand side in Eq. 3.6 is expanded in inverse powers of $R$. The dotted lines are $\mathcal{O}(R^{-5})$ NG terms which were left in the plot only to indicate the divergence of the expansion as the tachion singularity is approached.

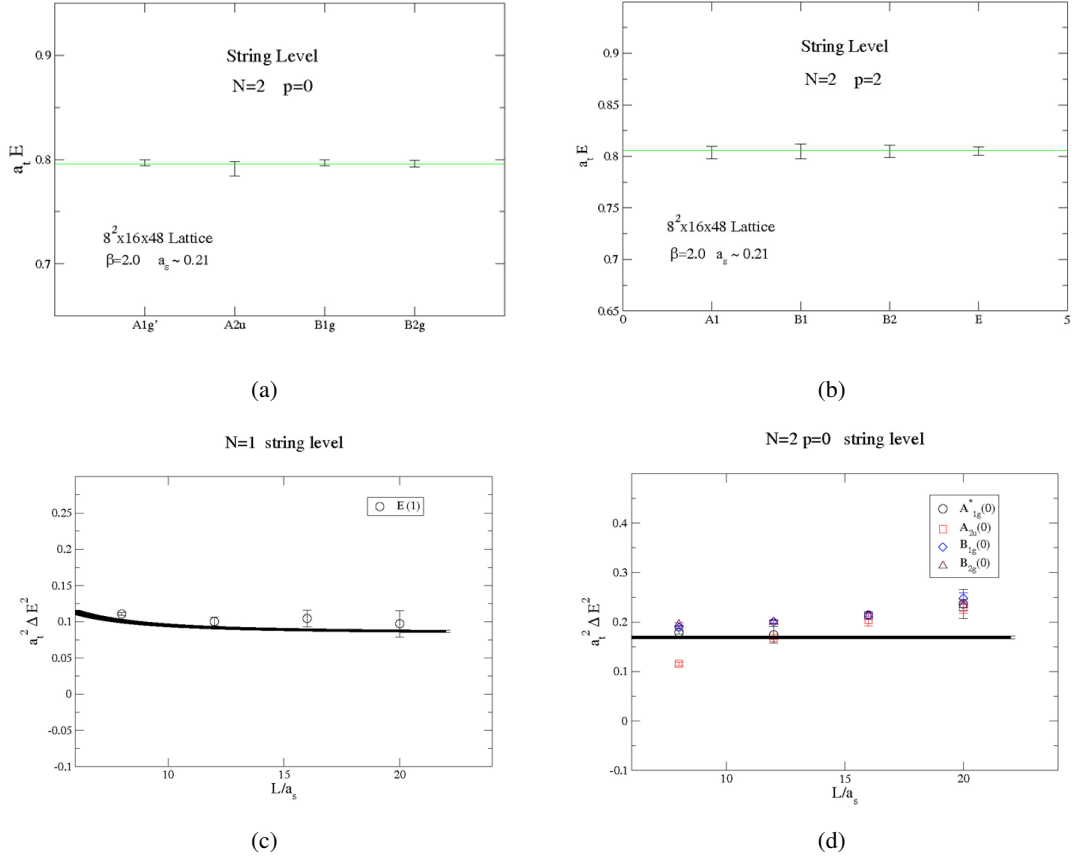

**Figure 15:** With data taken from [27], (a) and (b) illustrate excellent string degeneracies at fixed $R = aL$ for four different string states with quantum numbers which were discussed in Fig. 14b. In (c) and (d) torelon excitations are plotted against the predictions of Eq. 3.6 as L is varied. Deviations for large L in plot (d) which has several string states with the same string quantum numbers suggest that the lattice systematics will require improvements.

### 6.2 Torelon spectrum of the 4D SU(3) gauge model

The main author of the 4D SU(3) torelon excitation spectrum is Maresca with details of her work reviewed in [13, 27, 39]. The method is similar to the lattice technology which was used to obtain the Dirichlet string spectrum in section 5.3. To indicate the current status of the torelon spectra in the 4D SU(3) gauge model, selected energy levels are shown in Fig. 15. Given the difficulty of these calculations, the results are in reasonably good agreement with string theory expectations, but there are some distortions. Although the evidence for the torelon string spectrum in the 4D SU(3) gauge model is not as strong as the 3D Z(2) simulations, the full complex lattice technology is in place for extended and more comprehensive simulations which hopefully will be reported at Lattice 2006.

**Closing remarks and acknowledgements**

It was shown that the effective QCD string action can be probed now in a meaningful and well-defined framework to address important questions, pertinent to string theory. Appreciating the efforts of so many contributors, I would like to hope that the talk marks the end of the beginning in this exciting field. Finally, I thank the organizers of the conference at Trinity College for making Lattice 2005 stimulating and enjoyable.

## References

[1] H. B. Nielsen and P. Olesen, *Vortex line models for dual strings*, Nucl. Phys. B **61** (1973) 45.

[2] A. M. Polyakov, *Quark confinement and topology of gauge groups*, Nucl. Phys. B **120** (1977) 429.

[3] M. Lüscher, K. Symanzik, P. Weisz, *Anomalies of the free loop wave equation in the WKB approximation*, Nucl. Phys. B **173** (1980) 365.

[4] M. Lüscher, *Symmetry-breaking aspects of the roughening transition in gauge theories*, Nucl. Phys. B **180** (1981) 317.

[5] P. Hasenfratz, R.R. Horgan, J. Kuti, J.M. Richard, *The effects of colored glue in the qcd motivated bag of heavy quark–antiquark systems*, Phys. Lett. B **95** (1980) 299.

[6] S. Perantonis, C. Michael, *Static potentials and hybrid mesons from pure SU(3) lattice gauge theory*, Nucl. Phys. B **347** (1990) 854; I.J. Ford, R.H. Dalitz, J. Hoek, *Potentials in pure QCD on 32**4 lattices* Phys. Lett. B **208** (1988) 286.

[7] J. Polchinski and A. Strominger, *Effective string theory*, Phys. Rev. Lett. **67** (1991) 1681.

[8] K.J. Juge, J. Kuti, C.J. Morningstar, *Gluon excitations of the static quark potential and the hybrid quarkonium spectrum*, Nucl. Phys. Proc. Suppl. **63** (1998) 326.

[9] M.J. Teper, *SU(N) gauge theories in (2+1)-dimensions* Phys. Rev. D bf 59 (1999) 014512.

[10] M. Caselle, R. Fiore, F. Gliozzi, M. Hasenbusch, P. Provero, *String effects in the Wilson loop: a high precision numerical test*, Nucl. Phys. B **486** (1997) 245.

[11] P. Hoppe and G. Munster, *The interface tension of the three-dimensional Ising model in two loop order*, Phys. Lett. A **238** (1998) 265.

[12] K.J. Juge, J. Kuti, C.J. Morningstar, *QCD string formation and the Casimir energy*, Proceedings of *Color confinement and hadrons in quantum chromodynamics*, Wako, Japan (2004)233 [hep-lat/0401032].

[13] K.J. Juge, J. Kuti, F. Maresca, C.J. Morningstar, M. Peardon, *unpublished manuscript*.

[14] M. Billó, M. Caselle, M. Hasenbusch, M. Panero, *QCD string from D0 branes*, Proceedings of Lattice 2005, PoS(LAT2005)309.

[15] M.J. Teper, *Large N*, Proceedings of Lattice 2005, PoS(LAT2005)256 [hep-lat/0509019].

[16] R Narayanan, H. Neuberger, *Phases of planar QCD on the torus*, Proceedings of Lattice 2005, PoS(LAT2005)005 [hep-lat/0509014].

[17] R.C. Brower, *String/Gauge Duality: (re)discovering the QCD String in AdS Space*, Acta Phys. Polon. B **34** (2003) 5927 [hep-th/0508036].

[18] Z. Prkacin, G.S. Bali, T. Dussel, T. Lippert, H. Neff, K. Schilling, *Anatomy of String Breaking in QCD*, Proceedings of Lattice 2005, PoS(LAT2005)308.

[19] F. Gliozzi, *Screening of sources in higher representations of SU(N) gauge theories at zero and finite temperature*, Proceedings of Lattice 2005, PoS(LAT2005)196 [hep-lat/0509085].

[20] J.M. Kosterlitz and D.J. Thouless, *Ordering, metastability and phase transitions in two-dimensional systems*, J. Phys. C **6** (1973) 1181.

[21] H. van Beieren, *Exactly Solvable Model for the Roughening Transition of a Crystal Surface*, Phys. Rev. Lett. **38** (1977) 993.

[22] S. Flach and K. Kladko, *Perturbtion analysis of weakly discrete kinks*, Phys. Rev. E bf 54 (1995) 2912.

[23] A. M. Polyakov, *Fine structure of strings*, Nucl. Phys. B **268** (1986) 406.

[24] A. Herat, R. Rademacher, and P. Suranyi, *Curved, extended classical soutions: The undulating kink*, Phys. Rev. D bf 63 (2001) 027702.

[25] M. Lüscher and P. Weisz, *String excitation energies in SU(N) gauge theories beyond the free-string approximation*, JHEP **0407** (2004) 014.

[26] J.M. Drummond, *Universal subleading spectrum of effective string theory*, [hep-th/0411017]; J. Kuti, *unpublished*.

[27] F. Maresca, *Comparing the excitations of the periodic flux tube with effective string models*, Ph.D. Thesis, Trinity College, Dublin, 2004.

[28] J.F. Arvis, *The exact $q\bar{q}$ potential in Nambu string theory*, Phys. Lett. B **127** (1983) 106.

[29] M. Lüscher and P. Weisz, *Quark confinement and the bosonic string*, JHEP **0207** (2002) 049.

[30] M. Caselle, M. Pepe (Bern U.), A. Rago, *Static quark potential and effective string corrections in the (2+1)-d SU(2) Yang-Mills theory*, JHEP **0410** (2004) 005.

[31] P. Majumdar, *Continuum limit of the spectrum of the hadronic string*, [hep-lat/0406037].

[32] H. Meyer and M. Teper, *Confinement and the effective string theory in $SU(N \to \infty)$: a lattice study*, JHEP **0412** (2004) 031.

[33] T.T. Takahashi and H. Suganuma, *Detailed analysis of the gluonic excitation in the three-quark system in lattice QCD*, Phys. Rev. D bf 70 (2004) 074506.

[34] Ph. de Forcrand and O. Jahn, *The baryon static potential from lattice QCD*, Nucl. Phys. A **755** (2005) 475; *Baryons and confining strings*, Nucl. Phys. Proc. Suppl. **129** (2004) 700.

[35] K. Holland and J. Kuti, *unpublished*.

[36] K.J. Juge, J. Kuti, C.J. Morningstar, *Excitations of the static quark anti-quark system in several gauge theories*, Proceedings of *Color confinement and hadrons in quantum chromodynamics*, Wako, Japan (2004) 221 [hep-lat/0312019].

[37] K.J. Juge, J. Kuti, C.J. Morningstar, *The fine structure of the QCD string spectrum*, Phys. Rev. Lett. **90** (2003) 161601.

[38] G.S. Bali and A Pineda, *QCD phenomenology of static sources and gluonic excitations at short distances*, Phys. Rev. D bf 69 (2004) 094001.

[39] K.J. Juge, J. Kuti, F. Maresca, C.J. Morningstar, M. Peardon, *Excitations of torelon*, Nucl. Phys. Proc. Suppl. **129** (2004) 703.